\documentclass[fleqn,usenatbib]{mnras}

\usepackage{newtxtext,newtxmath}

\usepackage[T1]{fontenc}

\DeclareRobustCommand{\VAN}[3]{#2}
\let\VANthebibliography\thebibliography
\def\thebibliography{\DeclareRobustCommand{\VAN}[3]{##3}\VANthebibliography}

\usepackage{graphicx}	
\usepackage{amsmath}	
\usepackage{amssymb}	

\title[Proton Core Behaviour in Switchbacks]{Proton Core Behaviour Inside Magnetic Field Switchbacks}

\author[T. Woolley et al.]{Thomas Woolley,$^{1}$\thanks{E-mail: thomas.woolley15@imperial.ac.uk}
Lorenzo Matteini,$^{1}$
Timothy S. Horbury,$^{1}$
Stuart D. Bale,$^{2,3,4,1}$
Lloyd D. Woodham,$^{1}$
\newauthor
Ronan Laker,$^{1}$
Benjamin L. Alterman,$^{5}$
John W. Bonnell,$^{3}$
Anthony W. Case,$^{6}$
Justin C. Kasper,$^{7,6}$
\newauthor
Kristopher G. Klein,$^{8}$
Mihailo M. Martinovi\'c,$^{8,9}$
Michael Stevens$^{6}$ 
\\
$^{1}$ Department of Physics, Imperial College London, London SW7 2AZ, UK \\
$^{2}$ Physics Department, University of California, Berkeley, CA 94720-7300, USA \\
$^{3}$ Space Sciences Laboratory, University of California, Berkeley, CA 94720-7450, USA \\
$^{4}$ School of Physics and Astronomy, Queen Mary University of London, London E1 4NS, UK \\
$^{5}$ Space Science and Engineering, Southwest Research Institute, 6220 Culebra Road, San Antonio, TX 78238, USA \\
$^{6}$ Smithsonian Astrophysical Observatory, Cambridge, MA 02138 USA \\
$^{7}$ Climate and Space Sciences and Engineering, University of Michigan, Ann Arbor, MI 48109, USA \\
$^{8}$ Lunar and Planetary Laboratory, University of Arizona, Tucson, AZ 85721, USA \\
$^{9}$ LESIA, Observatoire de Paris, Universite PSL, CNRS, Sorbonne Universite, Universite de Paris, 5 place Jules Janssen, 92195 Meudon, France
}

\date{Accepted XXX. Received YYY; in original form ZZZ}

\pubyear{2020}

\begin{document}
\label{firstpage}
\pagerange{\pageref{firstpage}--\pageref{lastpage}}
\maketitle

\begin{abstract}
During Parker Solar Probe's first two orbits there are widespread observations of rapid magnetic field reversals known as switchbacks. These switchbacks are extensively found in the near-Sun solar wind, appear to occur in patches, and have possible links to various phenomena such as magnetic reconnection near the solar surface. As switchbacks are associated with faster plasma flows, we questioned whether they are hotter than the background plasma and whether the microphysics inside a switchback is different to its surroundings. We have studied the reduced distribution functions from the Solar Probe Cup instrument and considered time periods with markedly large angular deflections, to compare parallel temperatures inside and outside switchbacks. We have shown that the reduced distribution functions inside switchbacks are consistent with a rigid phase space rotation of the background plasma. As such, we conclude that the proton core parallel temperature is the same inside and outside of switchbacks, implying that a T-V relationship does not hold for the proton core parallel temperature inside magnetic field switchbacks. We further conclude that switchbacks are consistent with Alfv\'enic pulses travelling along open magnetic field lines. The origin of these pulses, however, remains unknown. We also found that there is no obvious link between radial Poynting flux and kinetic energy enhancements suggesting that the radial Poynting flux is not important for the dynamics of switchbacks.
\end{abstract}




\section{Introduction}
\label{intro}
Despite the prediction \citep{Parker_1958} and detection \citep{Gringauz_1960, Neugebauer_1962} of a supersonic wind from the Sun more than 60 years ago, it is still unknown how the thermal energy of the million-Kelvin corona is converted into the bulk kinetic energy of the solar wind flow. In situ plasma observations throughout the heliosphere reveal ubiquitous non-thermal particle velocity distribution functions (VDFs) in the plasma, suggesting that the heating and release of the solar wind close to the Sun, as well as its non-adiabatic expansion in interplanetary space are related to kinetic processes that regulate the energy exchanges between particles and fields (e.g. \citealt{Marsch_2006, Verscharen_2019}).

The solar wind is known to display a temperature-velocity (T-V) relationship on large scales over different streams (e.g. \citealt{Burlaga_1973, Lopez_1986}), but the exact drivers of this relationship remain unknown. Recent work suggests that the T-V relationship holds within a single stream \citep{Horbury_2018} and evolves as a function of distance \citep{PerroneD2019_TV}. This leads to questions about whether the T-V relationship also holds in individual small scale structures. To address this, it is important to consider VDFs in the solar wind which could carry fundamental information about the processes responsible for the heating and acceleration of the plasma close to the Sun. However, as solar wind expansion modifies and reprocesses distributions, measuring more pristine plasma conditions is fundamental to make a direct link between signatures observed in situ and processes at the Sun.

To this aim, the Parker Solar Probe (PSP) mission \citep{fox_2016} was designed to measure the young solar wind. During its first perihelion pass in November 2018 PSP reached a closest approach of 36.5 $R_{S}$. Prior to this, the closest in-situ measurements to the Sun were made by the Helios probes in the 1970s (62 $R_{S}$). An unexpected result from PSP’s first orbit was the detection of ubiquitous magnetic field reversals (switchbacks) \citep{Bale_2019_Nature} in the young solar wind, associated with intense enhancements of the flow velocity, up to twice the local Alfv\'en speed \citep{Kasper2019_Nature}.

These switchbacks are discrete, rapid and asymmetric magnetic field deflections away from the background field that can reverse the local magnetic field polarity. They have durations that last from a few seconds to tens of minutes, indicating that they are a multi-scale phenomenon \citep{Dudok_de_Wit_2020}. They also seem to occur in patches which are separated by regions of more quiet and stable radial magnetic field  \citep{Bale_2019_Nature, Horbury_2020}. Switchbacks are Alfv\'enic fluctuations with constant magnetic field intensity |B|, implying that the local plasma velocity inside a switchback is faster than the background flow \citep{Matteini_2014}. As a consequence, they also carry significantly larger momentum and kinetic energy than the surrounding plasma \citep{Horbury_2018}.

Magnetic switchbacks have also been observed beyond 64 $R_{S}$ \citep{Matteini_al_2015a, Horbury_2018}, although with different properties than in PSP data and mostly in the fast solar wind. Intriguingly, during its first perihelion, PSP was embedded in Alfv\'enic slow wind coming from a small coronal hole \citep{Bale_2019_Nature, Kasper2019_Nature}, similar to that previously discussed \citep{Damicis_Bruno_2015, Stansby2019, Perrone_al_2020}, revealing that these structures are more common than previously expected and suggesting that they could play a fundamental role in different types of streams and sources. 

Three questions about switchbacks then arise:
\begin{enumerate}
    \item Since the switchbacks are faster are they also hotter than the background plasma? This would be expected if typical solar wind T-V relationships are upheld inside these structures.
    \item  Is the plasma inside a switchback distinctly different from the background plasma? If it is then this would imply that switchbacks are a transient-like event from a source region that is distinct from that which generates the background plasma. On the other hand, if switchback and background plasma are similar then it is possible that switchbacks are local perturbations of the background plasma (e.g. a propagating non-linear Alfv\'enic pulse \citealt{Squire_2020}).
    \item Do switchbacks play a dynamical role in the generation of the solar wind? \citet{Mozer_2020} showed that switchbacks carry some significant radial Poynting flux that can eventually do work in accelerating the plasma; are the fastest switchbacks characterised by the largest Poynting flux?
\end{enumerate}

In this paper we address these questions by analysing magnetic field switchbacks which complete a full reversal in the radial component $B_{R}$, corresponding to the largest acceleration of the bulk plasma. These switchbacks provide the only opportunity to compare the parallel temperature inside and outside switchbacks using the Solar Probe Cup's radial measurements. We discuss the behaviour of the full ion VDF during the magnetic field rotation and highlight the thermodynamic properties of the proton core population. We also measure the radial Poynting flux's evolution within these structures and verify whether it is directly related to the plasma kinetic energy.

\section{Data}

In this work we used data from the FIELDS \citep{Bale_2016} and SWEAP \citep{Kasper_2016} instrument suites in the Radial Tangential Normal ($RTN$) coordinate system \citep{hapgood_1992}. We defined the magnetic field cone angle ($\theta_{BR}$) as the angle between the local magnetic field vector and the $R$ direction. It took a value between 0$^{\circ}$, when the magnetic field was exactly radial, and 180$^{\circ}$, when the magnetic field was exactly anti-radial.  .

\subsection{SWEAP}

The SWEAP instrument suite \citep{Kasper_2016} consists of two electron electrostatic analysers (SPAN-E; \citealt{Whittlesey_2020}), one ion electrostatic analyser (SPAN-I), and a Faraday cup (Solar Probe Cup; SPC; \citealt{Case_2020}). SPC and SPAN-I have complimentary fields of view. SPAN-I is situated on the ram side of the spacecraft behind the heat shield and measures a three-dimensional distribution function. SPC is radially orientated towards the Sun and measures a one-dimensional reduced ion distribution function, $F(v_{R})$, of the incoming solar wind that is blocked by the heat shield. 

The amount of solar wind measured by each instrument changes with the plasma flow relative to the spacecraft. For radial flows SPC is more appropriate whereas flows with a large -T velocity component relative to the spacecraft favour the use of SPAN-I. During the early phases of the mission, and at larger heliocentric distances, SPC is better suited for ion plasma measurements. As the spacecraft tangential velocity will increase for each subsequent perihelion pass, SPAN-I will capture more of the ion distribution in later encounters.

Here we processed the level 2 SPC-measured $F(v_{R})$ in accordance with the procedure outlined by \citet{Case_2020}. During PSP's first two perihelia, SPC's measurement cadence was typically between 1.1 and 4.6 samples/sec.

\subsection{FIELDS}

The FIELDS instrument suite \citep{Bale_2016} uses a variety of instruments to measure the magnetic and electric fields in the solar wind. These include two flux gate magnetometers (MAG), a search coil magnetometer (SCM), and five voltage probes. The magnetic field data used in this work was from the MAG instruments and was down sampled to the cadence of the SPC $F(v_{R})$.

We used 2 dimensional DC electric field data from four voltage probes approximately, but not exactly, in the T-N plane to calculate the radial Poynting flux ($S_{R} = \frac{1}{\mu_{0}}\left (\boldsymbol{E} \times \boldsymbol{B}  \right )\cdot \widehat{\boldsymbol{R}}$). Using these two components of E along with all three components of B allowed us to fully characterize the radial component of the Poynting flux. For the scales of interest, the planar electric field was dominated by the motional electric field ($\boldsymbol{E} = -\boldsymbol{v} \times \boldsymbol{B}$).



\section{Fitting the Proton Core}
\label{ProtonFit}
We manually chose switchbacks from PSP's first two perihelia that were some of the largest deflections with durations $>$5 minutes. For each case, we manually selected an interval that included both the switchback and background plasma for comparison.

We identified the proton core (pc) as the largest amplitude peak in each reduced distribution function. We fitted the following Gaussian to the seven points encompassing this peak (three adjacent points on either side):

\begin{equation}
 F_{pc}(v_{R}) = \frac{n_{pc}}{\sqrt{\pi }w_{R}}\cdot \exp\left (  \frac{(v_{R}-v_{pc, R})^{2}}{w_{R}^{2}}\right )
\end{equation}
with thermal speed:
\begin{equation}
w_{R}^{2} = \frac{2k_{B}T_{pc, R}}{m_{p}}.
\end{equation}
This was the proton core of the distribution function with a radial temperature ($T_{pc, R}$), number density ($n_{pc}$), and mean radial velocity ($v_{pc, R}$). Quality checks were then used to ensure a consistent level of fit.

We constrained the fits so that the density did not exceed $F(v_{R})$'s total density. We only kept fits for which the residuals were less than 0.05 cm$^{-3}$km $^{-1}$s. This filtered out cases which did not accurately represent the data points. Despite these filters, the core temperature was overestimated in some of the $F(v_{R})$ with large proton beams (e.g. $n_{beam}/n_{pc} \sim 1$), therefore we removed these $F(v_{R})$ manually. Note that \citet{Verniero_2020} has studied large proton beams in more detail.  We validated our fits by comparing the extracted physical quantities to those in the level 3 SPC data files. As our fitting method was similar, we assign the same uncertainties on fitted parameters as presented by \citet{Case_2020}. These are 9\%, 3\% and 19\% for the density, radial velocity and temperature respectively.

\begin{figure}
  \includegraphics[width=\columnwidth]{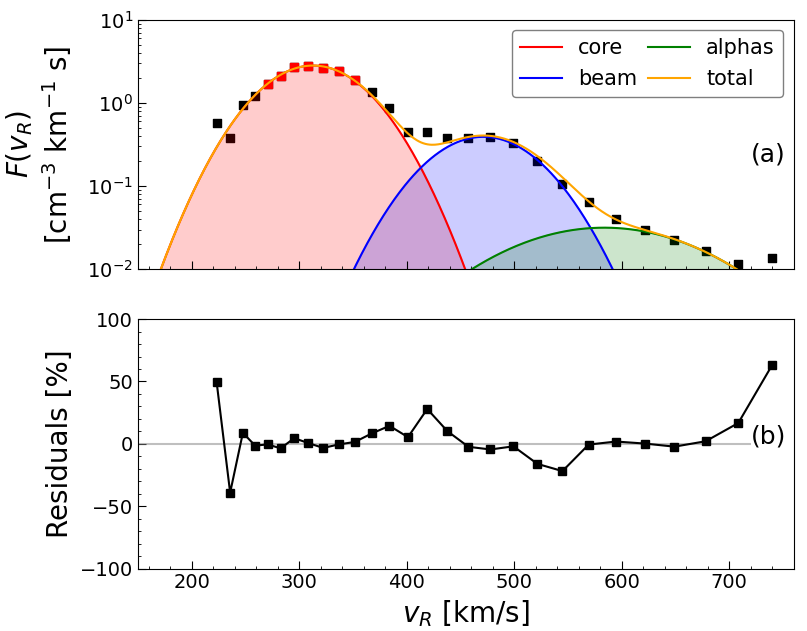}
 \caption[]{\small Fits of the proton core (red), proton beam (blue) and alpha (green) populations along with SPC's reduced distribution function (black data points) within the background plasma (5th November 2018 01:22:45). Panel (a) shows the reduced distribution function and population fits. The red points indicate the data used to fit the core population and the orange line shows the sum of the three population fits. Panel (b) shows the residuals in units of percent $\left( 100 \times \frac{\mathrm{data}-\mathrm{fit}}{\mathrm{data}} \right)$.}
 \label{figcorefits}
 \end{figure}

In the background plasma, it was occasionally possible to also fit the beam and alpha populations using a similar procedure as that applied to the core. Fig.~\ref{figcorefits} shows an example fit to the proton core (red), proton beam (blue) and alpha (green) populations in the background plasma near to perihelion. Panel (a) shows the $F(v_{R})$ as measured by SPC (black data points) and the fit populations. The sum of the distributions (orange line) closely follows the data points, suggesting that the extracted physical quantities represent the measurements well. Panel (b) shows residuals in units of percent, confirming the fit quality. Note that the alpha particles' velocity was shifted by a factor of $\sqrt{2}$ (to $\sim$ 600 km s$^{-1}$) in SPC because of the alpha's energy-to-charge ratio. Accounting for this shift, the alphas are found to travel at approximately the local Alfv\'en speed and have a higher $v_{R}$ than the proton core. 

In the cases we selected, the proton beam and alphas were sufficiently well separated from the core that fitting them did not impact our proton core fits. We only mentioned them in this section for completeness.

\section{Multi-species Motion under Alfv\'enic Fluctuations}
\label{background}

In order to make use of the SPC measurements, it was important to understand how different plasma species behaved inside and outside magnetic field switchbacks. Outside of switchbacks, the background magnetic field was approximately anti-radial close to perihelion \citep{Bale_2019_Nature, Kasper2019_Nature}. As a result, SPC's radial temperature measurements corresponded to the component parallel to the local magnetic field.

When PSP observed an Alfv\'enic fluctuation with $\left | B \right |$ constant, the highly correlated velocity and magnetic field caused the ion VDF to rotate in phase space around the velocity corresponding to the local wave speed in the plasma reference frame. As a result, SPC measured the VDF at different angles to the magnetic field. As protons have anisotropic temperatures with respect to the local magnetic field in the solar wind, SPC measured a different radial temperature as a function of $\theta_{BR}$. This is why direct comparisons of SPC cuts and parallel temperatures inside and outside switchbacks were only possible for full reversals of the local magnetic field.

\begin{figure}
  \includegraphics[width=\columnwidth]{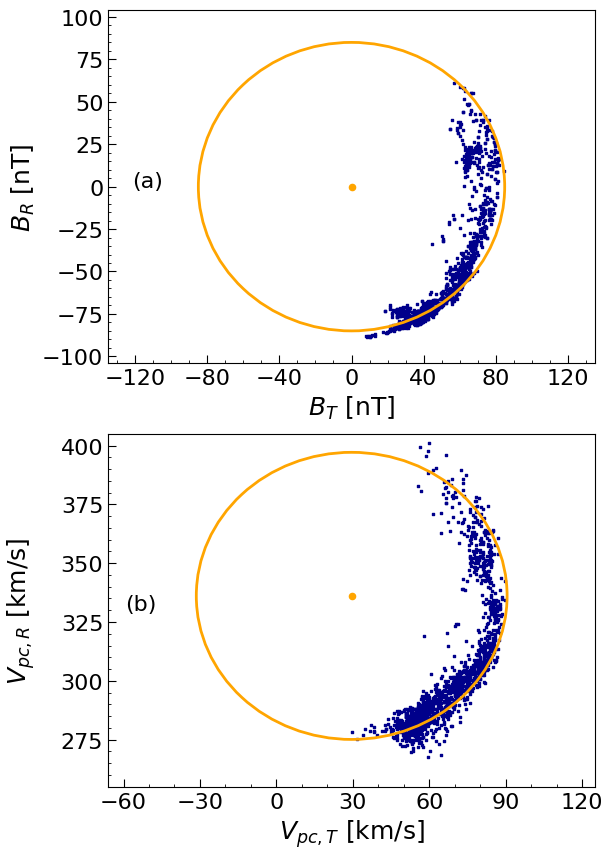}
 \caption[]{\small Magnetic field and proton core velocity in the T-R plane from an $\approx$ 25 minute interval (5th November 2018 06:55:05 - 07:19:30) close to perihelion. Panel (a) shows the magnetic field (dark blue points) with a circle of radius 82 nT ($\simeq \left \langle \left | B \right | \right \rangle$) centred on (0, 0) nT over plotted in orange. Similarly,  panel (b) shows the velocity of the proton core (dark blue points) with a circle of radius 60 km s$^{-1}$ ($\simeq v_{phase}$) over plotted in orange. The centre of the velocity circle is (29, 342) km s$^{-1}$.}
 \label{figBVfluct}
 \end{figure}

Fig.~\ref{figBVfluct} shows the magnetic field and proton core velocity during Alfv\'enic fluctuations for a chosen interval. As these were constant $\left | B \right |$ structures, the magnetic field vector was confined to move on the surface of a sphere with constant radius equal to $\left | B \right |$ centred at (0, 0, 0) nT. When this motion was projected into a plane, it appeared as an arc as shown in panel (a). Similarly, the velocity vector was confined to the surface of a sphere in velocity space, the projection of which is shown in panel (b). This shows that the velocity fluctuations were also rotations (i.e. constant magnitude) in a reference frame that was close to the local wave speed, which is typically the Alfv\'en speed \citep{Matteini_al_2015a}. As such, the velocity sphere's radius was $\simeq v_{phase}$. The centre of the velocity sphere was approximately the local de Hoffman Teller frame, i.e. the frame in which the motional electric field associated with the fluctuations vanished.

A further consequence of the phase space rotation discussed above was that every particle above (below) the local wave speed travelled slower (faster) within a switchback, while the particles that streamed at exactly the local wave speed were neither accelerated nor decelerated during switchbacks \citep{Matteini_al_2015a}. In general, the local wave speed usually sits somewhere between the proton core velocity and the proton beam velocity. As such, the proton core accelerates while the proton beam decelerates during switchbacks \citep{Neugebauer_2013}. When the magnetic field is approximately perpendicular to the radial direction, the proton core and beam have the same radial velocity and hence the two populations overlap. This makes it difficult to distinguish the two populations in SPC data. As the alpha particles appear to stream at the local wave speed close to the Sun (Fig.~\ref{figcorefits}), they are expected to remain at a constant velocity during switchbacks.

 \begin{figure*}
 \centering
  \includegraphics[width=2.0\columnwidth]{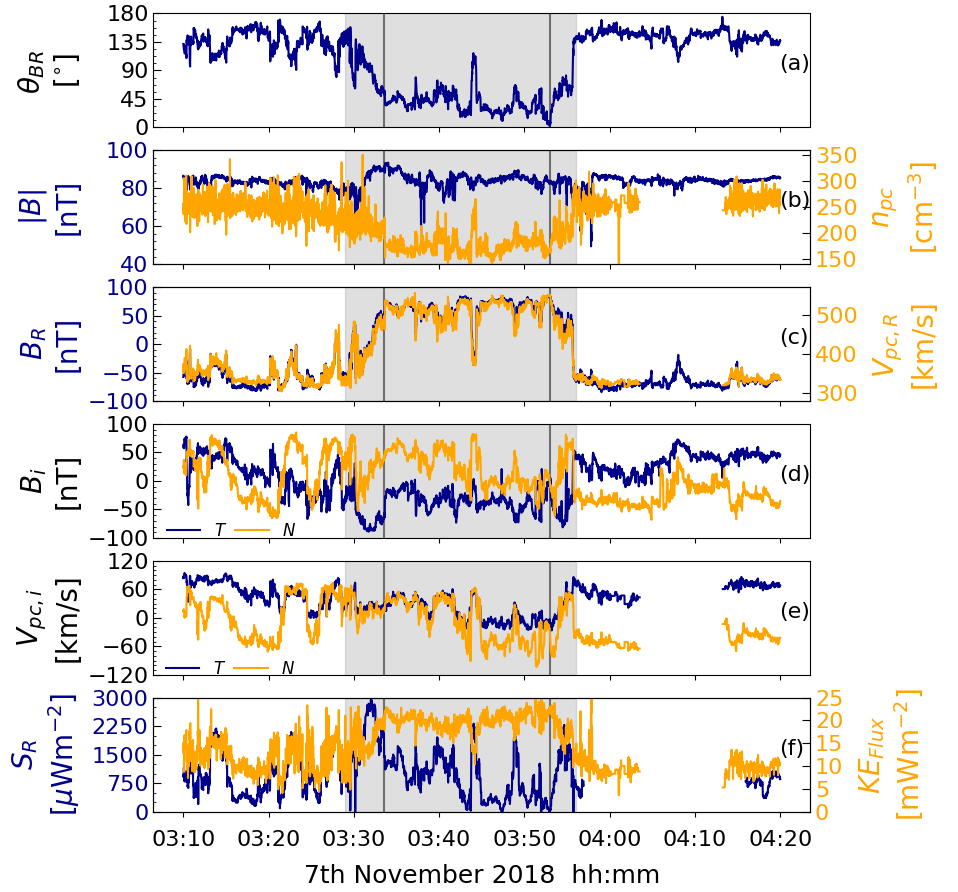}
 \caption[]{\small A selected switchback showing: (a) the magnetic field cone angle $\theta_{BR}$, (b) the magnitude of the magnetic field $ \left | B \right | $ and the proton core density $n_{pc}$, (c) the radial components of the magnetic field $B_{R}$ and the proton core velocity $v_{pc, R}$, (d) the tangential and normal components of the magnetic field $B_{T}$ and $B_{N}$, (e) the tangential and normal components of the proton core velocity $v_{pc, T}$ and $v_{pc, N}$, (f) the radial Poynting flux $S_{R}$ and the proton core kinetic energy flux. The grey shaded region highlights the magnetic field switchback and the thin vertical lines indicate the inner region with parallel magnetic field.}
 \label{figcase1}
 \end{figure*}
 
\section{Results}
\subsection{Case Study: 7th November 2018}
We focused on a specific switchback that occurred on the 7th November 2018 when PSP was approximately  37.6 $R_{S}$ from the Sun (see Fig.~\ref{figcase1}). This switchback lasted for approximately 25 minutes and panel (a) shows that $\theta_{BR}$ was close to 0$^{\circ}$ for most of this time. The magnetic field magnitude in the switchback remained approximately constant at a slightly greater value than the background field with occasional short-lived fluctuations. This magnitude increase was counteracted by a density decrease in the switchback plasma to maintain a similar total pressure to the background plasma (panel (b)). Panel (c) shows that this switchback started with a slow deflection away from the background field and ended with a sharp, rapid return to the background orientation. In the middle of the switchback at approximately 03:44 there was a small sharp feature which caused the magnetic field to almost return to the background orientation briefly. The radial velocity profile (orange line) followed $B_{R}$ closely, as the plasma fluctuations were Alfv\'enic. Panel (d) highlights that the deflection of the magnetic field occurred in the negative $T$ direction with the first half of the deflection rotating towards the positive $N$ direction. It also indicates that the plasma fluctuations were Alfv\'enic when the magnetic field components are compared to the plasma proton core velocity components in panel (e). Finally, panel (f) shows both the radial Poynting flux (calculated from $\frac{1}{\mu_{0}} \mathbf{E} \times \mathbf{B}$) and the kinetic energy flux of the proton core population (Sect. \ref{PoyntingFlux}). There was a $\sim$ 20 minute interval without electric field or plasma data after the switchback which can be seen as the gap in the data. All the velocities that are plotted in Fig.~\ref{figcase1} are for the proton core population and come from the fitting procedure outlined in Sect. \ref{ProtonFit}.

\subsubsection{Temperatures}
\label{results_temp}
\begin{figure}
  \includegraphics[width=\columnwidth]{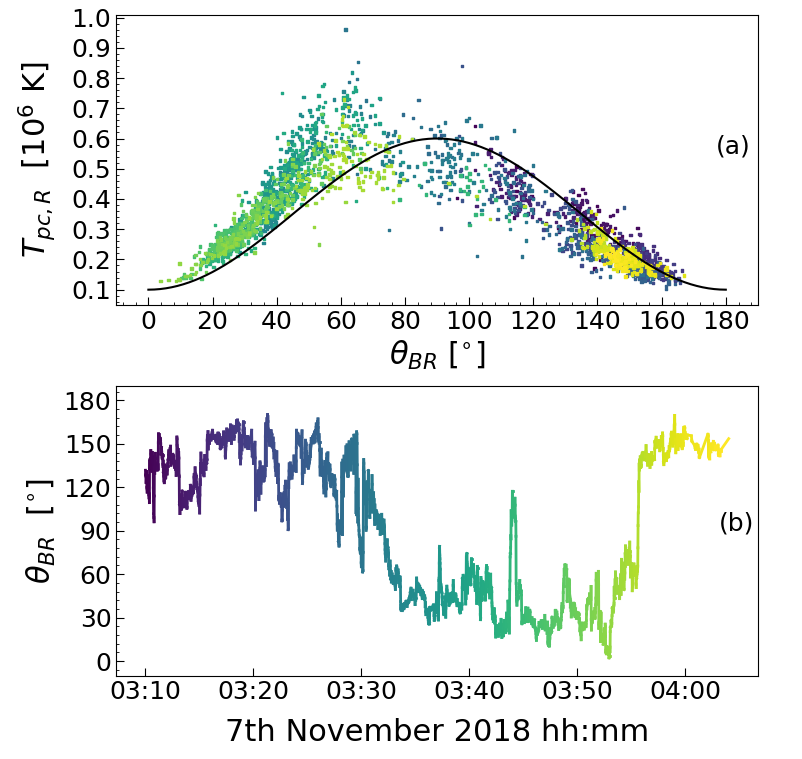}
 \caption[]{\small Proton core radial temperature within a single switchback on the 7th November 2018. Panel (a) shows the proton core radial temperature plotted against the magnetic field cone angle $\theta_{BR}$. The solid black line shows the expected response of the temperature (Eq. \ref{eq:temp}). Panel (b) shows a timeseries plot of $\theta_{BR}$. Both panels follow the same colour convention. The purple data points are the earliest and the yellow data points are the latest.}
 \label{fig:case1temp}
 \end{figure}

The proton core radial temperature ($T_{pc, R}$) for this case study is plotted in Fig.~\ref{fig:case1temp}. Panel (a) shows its dependence on magnetic field cone angle and panel (b) shows how the cone angle changed throughout the interval. The data points are coloured based on time: purple indicates the earliest times and yellow indicates the latest.

Panel (a) shows that $T_{pc, R}$ generally follows a geometrical prediction (black line) for an anisotropic core plasma with $T_{pc, \parallel} = $ 0.1$\times$ 10$^{\text{6}}$ K and $T_{pc, \perp} = $ 0.6$\times$ 10$^{\text{6}}$ K \citep{Kasper_et_al_2002, huang_2020}: \begin{equation}
T_{pc, R} = T_{pc, \parallel }\cos^{2}\theta_{BR}  + T_{pc, \perp}\sin ^{2}\theta_{BR}.
\label{eq:temp}
\end{equation}
This is consistent with the same anisotropic ($T_{pc, \perp} > T_{pc, \parallel}$) VDF being seen from different angles as the magnetic field orientation changes. This suggests that changes in $T_{pc, R}$ were due to changes in the geometrical cut through the VDF rather than variations in the underlying distribution. Panel (a) also shows that the proton core's parallel temperature was the same at 0$^{\circ}$ (within the switchback) and at 180$^{\circ}$ (in the background plasma). There are, however, some deviations from the geometrical prediction, which could be related to systematic errors in the SPC measurements. These will be addressed in future studies.

\subsubsection{Velocity Distribution Functions}

\begin{figure}
  \includegraphics[width=\columnwidth]{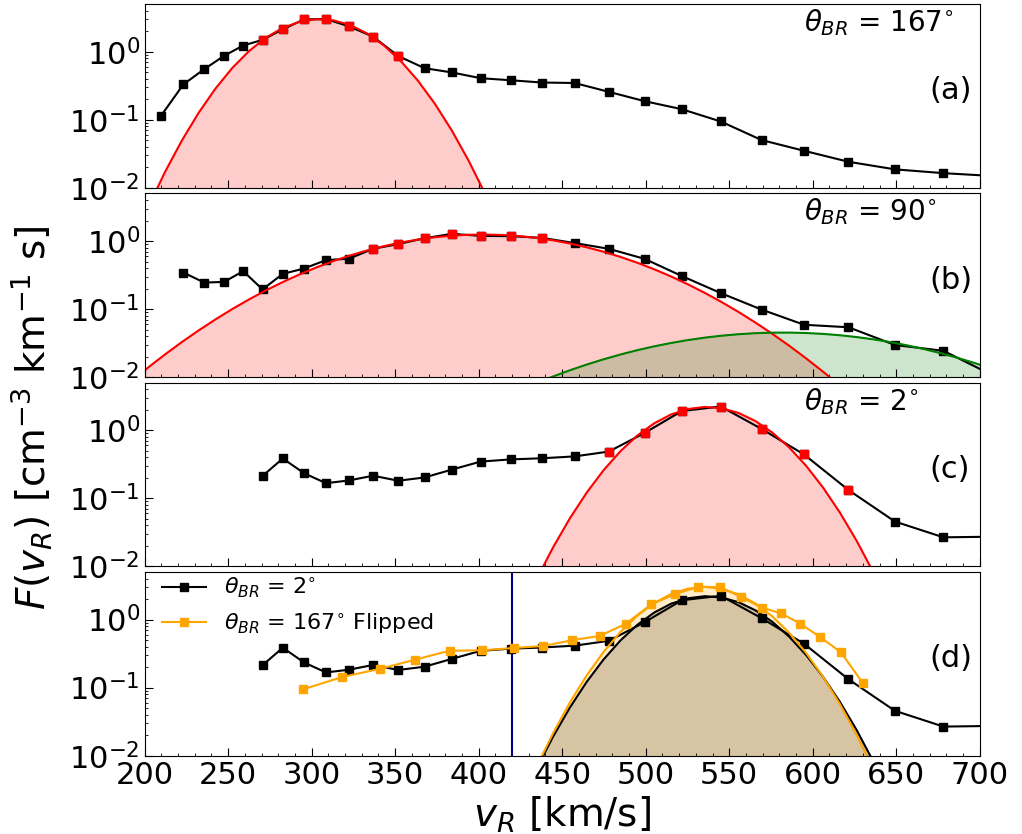}
 \caption[]{\small Reduced distribution functions measured by SPC for the 7th November switchback. Panels (a) – (c) show the SPC reduced distribution functions when $\theta_{BR}$ was 167$^{\circ}$ (03:25:59), 90$^{\circ}$ (03:29:47) and 2$^{\circ}$ (03:52:58). As such, panels (a) and (c) were obtained when the magnetic field was almost radial and anti-radial, whereas panel (b) was from a time during the switchback when the field was almost perpendicular to the radial direction. The red shaded areas show the proton core and the red points show the data used to fit the proton core. The green shaded area in panel (b) shows the fitted alpha particles. Panel (d) shows the distribution from panel (c) in black with the distribution from panel (a) rotated around $v_{pc, R} + V_{A}$ and over plotted in orange. The blue vertical line indicates the velocity around which the distribution from panel (a) was rotated.}
 \label{fig:case1distributions}
 \end{figure}

Fig.~\ref{fig:case1distributions} panels (a) through (c) present example $F(v_{R})$ from the approximately anti-parallel, perpendicular, and parallel cases. The red shaded region shows the core proton population. The green shaded region shows the alpha particle population which we could only estimate for the perpendicular field case. 

Panel (b) shows a case where the proton core and beam overlapped for $\theta_{BR}$ around 90$^{\circ}$. This was a consequence of both populations having the same radial speed and resulted in the core population obscuring the beam as discussed in Sect. \ref{background}. The temperature of this core distribution is larger than the temperature of either of the distributions in panels (a) and (c). It should be emphasised that this temperature difference is a direct result of the anisotropy of the plasma and not because of the core-beam overlapping. 

Panels (a) and (c) show the anti-parallel (background) and parallel $F(v_{R})$ respectively. As expected, the average velocity of the core population in the parallel case was higher due to the motion of the populations under Alfv\'enic fluctuations (see Sect. \ref{background}). This motion not only supported the idea that VDFs undergo a rigid phase space rotation as presented previously but also allowed an independent estimate of the local phase speed ($ v_{phase}$). It is worth noting that $ v_{phase}$ corresponds to the speed that fluctuations propagate in the plasma frame and hence, the frame in which the motional electric field of the fluctuations vanishes \citep{Matteini_al_2015a, Horbury_2020}. $ v_{phase}$ was estimated as 115 km s$^{-1}$ by considering the motion of the core population in the two $F(v_{R})$. This was consistent with the phase speed ($\sim $ 110 km s$^{-1}$) obtained from the correlation between $v_{pc, N}$ and $B_{N}$ fluctuations before the switchback and was in good agreement with the local Alfv\'en speed ($V_{A} \sim $ 110 km s$^{-1}$).

Panel (d) compares the $F(v_{R})$ from panel (a) rotated around a velocity $ v_{phase}$ ahead of the core population (blue line) and the $F(v_{R})$ from panel (c). The two distributions are very similar which is again consistent with a phase space rotation of the same VDF.

The average velocity of the alpha population was 412 km s$^{-1}$ which, due to the energy-to-charge ratio of alpha particles, was shifted by a factor $\sqrt{\text{2}}$ to $\sim$ 583 km s$^{-1}$ in SPC's $F(v_{R})$ (panel (b), Fig.~\ref{fig:case1distributions}). This average velocity was consistent with the alpha particles streaming faster than the proton core by $ v_{phase}$. As such, the alphas were located at the centre of the phase space rotation and hence remained at the same velocity for all magnetic field angles. We conclude that the tail seen above 600 km s$^{-1}$ in each $F(v_{R})$ was alpha particles.

\subsubsection{Poynting Flux}
\label{PoyntingFlux}
Panel (f) in Fig.~\ref{figcase1} shows the radial Poynting flux ($S_{R} = \frac{1}{\mu_{0}}\left (\boldsymbol{E} \times \boldsymbol{B}  \right )\cdot \widehat{\boldsymbol{R}}$) and the proton core kinetic energy flux through the period of study. The kinetic energy flux increased within the switchback because the core velocity increased. As such, the kinetic energy flux's profile (orange line) was very similar to the profile of the radial magnetic field and core velocity components.

The radial Poynting flux was small in the background plasma before the switchback but as the field began to rotate from sunward to anti-sunward polarity, this flux increased and reached a maximum at around $\theta_{BR}$ = 90$^{\circ}$. It then fell towards the background level as $B_{R}$ increased to its maximum value within the switchback. During the switchback, when $B$ was mainly radial, the Poynting flux was similar to the background level even though the velocity of the proton core was much higher. At the end of the switchback, when the field began to return to the background orientation, the Poynting flux once again increased. It reached a maximum around $\theta_{BR}$ = 90$^{\circ}$ before decreasing as the magnetic field returned to its pre-switchback orientation.

The proton core kinetic energy flux was always significantly larger than the radial Poynting flux (Fig.~\ref{figcase1}). The ratio of Poynting flux to kinetic energy flux was $\sim$1/20 in the background, when the field was typically close to anti-radial. At $\theta_{BR}$ = 90$^{\circ}$ when the radial Poynting flux was maximum, the ratio was also maximum ($\sim$1/5). However, since the amount of radial Poynting flux was similar to the background value and the kinetic energy was enhanced with respect to the background, the ratio inside the switchback was minimum ($\sim$1/40) when the field was nearly radial.

We obtained an expression for the ratio by noting that $v_{pc, R}  \gg v_{pc, T}, v_{pc, N}$ and $E \simeq -v_{pc} \times B$ for the study period (for brevity the subscript pc has been dropped for terms in the following equations):

\begin{equation}
\Gamma = \frac{S_{r}}{KE_{flux}} \sim \frac{\frac{1}{\mu_{0}}v_{R}B_{\perp}^{2}}{\frac{1}{2}\rho v_{R}^{3}}  \sim \frac{2}{v_{R}^{2}} \frac{B_{\perp}^{2}}{\mu_{0}\rho }.
\end{equation}
For spherically polarised fluctuations with approximately constant $\left | B \right |$ (see Fig.~\ref{figBVfluct}), we expect $\Gamma$ to be maximum at $\theta_{BR}$ = 90$^{\circ}$. This is because the increase of $v_{R}$ within the switchback is counteracted by the decrease of $B_{\perp}$ for $\theta_{BR}$ < 90$^{\circ}$. As the maximum occurs when $B_{R} \sim 0$ and $B_{\perp} \simeq B$, the maximum ratio can be written as:

\begin{equation}
\label{equ:maxratio}
\Gamma_{max} \sim 2  \left (\frac{V_{A}}{v_{R}}    \right )^{2}  \vline_{\theta_{BR}=90^{\circ}}.
\end{equation}
This provides an approximate expression for the upper limit of the radial Poynting flux energy contribution for any switchback. Eq. \ref{equ:maxratio} predicts a maximum ratio of $\sim$1/5 for the 7th November 2018 switchback which is consistent with our observations.

\subsection{Other Switchbacks}
\label{other_switchbacks}
To validate our findings, we investigated other switchbacks from PSP's first and second near-sun encounters (see Appendix~\ref{appendix} for times of switchbacks). They showed properties similar to the example presented above. For example, the 5th November 2018 switchback at 02:20 shown by \citet{Bale_2019_Nature} displayed similar Poynting and kinetic energy flux behaviour.

In order to obtain a meaningful sample, we relaxed our full magnetic field reversal requirement. Instead of selecting switchbacks that under went a full reversal, any switchback that had $>$ 8 data points with $\theta_{BR} <$ 30$^{\circ}$ were chosen. We fit the data for $\theta_{BR} <$ 30$^{\circ}$ and $\theta_{BR} >$ 150$^{\circ}$ separately using Eq. \ref{eq:temp} to get estimates for $T_{pc, \parallel}$ inside and outside (background) of a switchback respectively. Even with this relaxed condition, the number of switchbacks for which the inside and outside fits were successful was only 5, as most events did not rotate to $\theta_{BR} <$ 30$^{\circ}$.

\begin{figure}
  \includegraphics[width=\columnwidth]{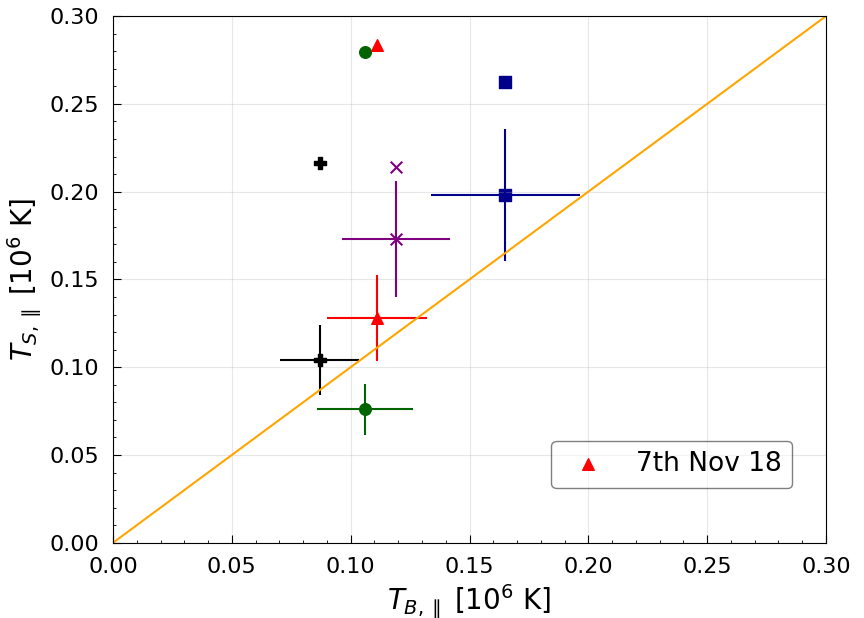}
 \caption[]{\small Proton core parallel temperatures inside ($T_{S, \parallel}$) and outside ($T_{B, \parallel}$) of switchbacks. There are two data points for each switchback which are indicated with the same marker and colour. The data with error bars are the proton core parallel temperatures obtained using the procedure in Sect.~\ref{other_switchbacks}. The data without error bars are the proton core parallel temperature predictions that arise by using the estimated T-V relationship at 35 $R_{S}$. The 7th November 2018 case study is shown with the red triangle. The orange line indicates $T_{S, \parallel} = T_{B, \parallel}$.}
 \label{fig:allcases}
 \end{figure}

Fig.~\ref{fig:allcases} shows $T_{pc, \parallel}$ in a switchback ($T_{S, \parallel}$) against $T_{pc, \parallel}$ in the background plasma ($T_{B, \parallel}$) and the orange line indicates where the $T_{S, \parallel}$ to $T_{B, \parallel}$ ratio is 1. There are two data points for each switchback which are indicated by the same colour and symbol. First, the points with error bars are the proton core parallel temperatures calculated from the procedure outlined above. Since we use an analogous fitting procedure we use a temperature uncertainty of 19\% as in \citet{Case_2020}. Second, the points without error bars are estimates of the proton core parallel temperature inside each switchback, which we obtained by using the T-V relationship at 35 $R_{S}$ \citep{PerroneD2019_TV}. The ratio of the switchback to background proton core parallel temperature for each switchback deviates strongly from the T-V prediction and is instead remarkably close to 1. This is consistent with $T_{S, \parallel}$ being the same as $T_{B, \parallel}$ and unrelated to the plasma velocity.

\section{Discussion}

The proton core parallel temperatures inside and outside of our case study (Fig.~\ref{fig:case1temp}) and other switchbacks (Fig.~\ref{fig:allcases}) indicate that the plasma is not hotter within full or near-full switchbacks. This suggests that the typical solar wind T-V relationship does not apply to the proton core parallel temperature inside magnetic field switchbacks and answers question (i) from Sect.~\ref{intro}. The $F(v_{R})$ measured before and during the 7th November 2018 switchback (Fig.~\ref{fig:case1distributions}) are consistent with a rigid rotation in velocity space \citep{Matteini_al_2015a}, leading to a core-beam swap inside the switchback \citep{Neugebauer_2013}. The centre of this velocity space rotation empirically agrees with the local phase velocity of fluctuations. Our results suggest that the plasma inside a switchback is not distinctly different from the background plasma (question (ii) in Sect.~\ref{intro}).

This seems to rule out that these structures are remnants of faster and hotter plasma directly injected in the corona that propagate through a slower background. Our findings support the idea that magnetic field switchbacks are large amplitude non-linear Alfv\'en waves propagating along open field lines such as those discussed by \citet{Gosling_2011}. It is not obvious how such structures can remain stable for prolonged times \citep{Landi_2006} but it seems that a constant field magnitude is required \citep{tenerani_2020}. Alfv\'en pulses could originate at the Sun through interchange reconnection events and propagate to large distances \citep{karpen_2017, Roberts_2018, sterling_2020}. Alternatively, they could also originate from the non-linear evolution of large amplitude fluctuations during expansion \citep{Squire_2020}.

The radial Poynting flux's observed profile (Panel (f), Fig.~\ref{figcase1}) is consistent with it being dominated by the electric field term $\sim v_{R} B_{\perp}$. This term is largest when $B$ is away from the radial, while it is small for radial or anti-radial field. Our observations are in agreement with the functional form presented in \citet{Mozer_2020} and suggest that the intermediate velocity switchbacks carry the largest amount of wave energy radially outwards, while the very fastest (i.e. full reversals) only carry small amounts (question (iii) in Sect.~\ref{intro}).

The ratio of radial Poynting to kinetic energy flux is consistent with negligible radial Poynting flux in the fastest switchbacks which suggests that the dynamics are not driven by the wave energy that switchbacks carry. We predict that the maximum ratio of radial Poynting to kinetic energy flux inside any switchback is given by Eq. \ref{equ:maxratio}. The ratio in a switchback will tend to the upper limit given by Eq. \ref{equ:maxratio} as $\theta_{BR}$ approaches 90$^{\circ}$, but will be considerably less elsewhere.

\section{Conclusion}

During the first perihelion pass of PSP ubiquitous local magnetic field reversals (switchbacks) were measured in the young solar wind \citep{Bale_2019_Nature, Kasper2019_Nature}. These switchbacks, which were associated with large increases in the plasma flow velocity, were very different to the switchbacks observed previously in Helios data \citep{Horbury_2018} and were resolved in more detail with the improved cadence of instruments on-board PSP. Here we have used a detailed analysis of ion distributions to address the microphysics inside switchbacks.

Our analysis suggests that the plasma inside switchbacks is not distinctly different from the background plasma. We have also shown that the proton core parallel temperature is not related to the enhanced velocity of switchbacks. These results are consistent with a phase space rotation of the plasma VDF, which could be a result of non-linear Alfv\'en pulses propagating through the background plasma. The origin and mechanisms that produce such Alfv\'enic pulses are still unknown.

We considered the behaviour of the radial Poynting flux and conclude that the fastest switchbacks do not carry the largest radial Poynting flux. Instead it is the intermediate velocity switchbacks and field rotations, where $B_R\sim0$, that transport the most wave energy radially outwards. This behaviour is what was expected from purely geometrical considerations about the motional electric field $\mathbf{E} = -\mathbf{v} \times \mathbf{B}$. We conclude that there is no obvious link between kinetic energy enhancement and radial Poynting flux in the largest switchbacks.

As a word of caution, in order to exploit the capabilities of the SPC sensor we could only investigate the largest switchbacks. As such, we cannot make general assumptions about all switchbacks from the case studies addressed here. From our work, the proton parallel temperature remains the same inside and outside of the largest switchbacks but previous work at 1 au \citep{woodham_2020} suggests that the temperature anisotropy and parallel temperature of the proton core depend on the magnetic field cone angle. Future work should address this using the SPAN instruments to determine whether the proton parallel temperature is the same inside intermediate velocity switchbacks.

Further work could also include a similar analysis for proton beams and alphas, but instrument limitations may make this study difficult. Instead, combining the data from the ion electrostatic analyser (SPAN) with that of SPC will allow a more comprehensive, 3-dimensional distribution function to be constructed. With a 3-dimensional distribution, temperature anisotropies of each species, for example, could be investigated in-depth. Solar Orbiter's recent launch presents the possibility of comparing measurements from both spacecraft. This will be especially helpful if the two spacecraft are connected to the same solar source region and may allow the radial and latitudinal evolution of the plasma to be studied in detail.

\section*{Acknowledgements}

TW was supported by STFC grant ST/N504336/1, TSH by STFC ST/S000364/1, RL by an Imperial College President's scholarship and LDW by ST/S000364/1. SDB acknowledges the support of the Leverhulme Trust Visiting Professor program. The SWEAP Team acknowledges support from NASA contract NNN06AA01C. KGK was supported by NASA grant 80NSSC19K0912.

\section*{Data Availability}

The data used in this research is all publicly available at: \url{https://cdaweb.gsfc.nasa.gov/index.html/}



\bibliographystyle{mnras}
\bibliography{references} 



\newpage
\appendix
\section{Switchback Times}
\label{appendix}
The times of the switchbacks in Sect.~\ref{other_switchbacks} are summarised below.

\begin{table}
\begin{tabular}{lll}
\hline
Time        & $T_{B, \parallel}$ [10$^{6}$K] & $T_{S, \parallel}$ [10$^{6}$K] \\ \hline
1 Nov 18 \quad 01:00 - 02:30  & 0.165 $\pm$  0.031      & 0.198 $\pm$  0.038 \\ 
5 Nov 18 \quad 13:45 - 14:20  & 0.106 $\pm$  0.020      & 0.076 $\pm$  0.014 \\
7 Nov 18 \quad 03:10 - 04:05  & 0.111 $\pm$  0.021      & 0.128 $\pm$  0.024 \\ 
31 Mar 19 \quad 10:24 - 10:40 & 0.087 $\pm$  0.017      & 0.104 $\pm$  0.020 \\ 
1 Apr 19 \quad 09:50 - 10:46  & 0.119 $\pm$  0.023      & 0.173 $\pm$  0.033 \\ \hline
\end{tabular}
\caption{Switchback times for cases presented in Fig.~\ref{fig:allcases}.}
\end{table}

\bsp	
\label{lastpage}
\end{document}